\pdfoutput=1
\documentclass[%
 reprint,
 amsmath,amssymb,longbibliography,
 aps,
 prb,
]{revtex4-2}
\usepackage{hyperref}

\usepackage{graphicx}
\usepackage{amsmath}
\usepackage{color}

\begin{document}

\title{Orbital-dependent self-energy effects and consequences for the superconducting gap structure in multi-orbital correlated electron systems}

\author{Kristofer Bj\"{o}rnson$^{1,2}$, Andreas Kreisel$^3$, Astrid T. R\o mer$^1$, Brian M. Andersen$^1$}
\affiliation{%
$^1$Niels Bohr Institute, University of Copenhagen, Jagtvej 128, DK-2200 Copenhagen, Denmark\\
$^2$Department of Physics and Astronomy, Uppsala University, Box 516, S-751 20 Uppsala, Sweden\\
$^3$Institut f\" ur Theoretische Physik, Universit\"at Leipzig, D-04103 Leipzig, Germany}

\date{\today}

\begin{abstract}
We perform a theoretical study of the effects of electronic correlations on the superconducting gap structure of multi-band superconductors. In particular, by comparing standard RPA-based spin-fluctuation mediated gap structures to those obtained within the FLEX formalism for an iron-based superconductor, we obtain directly the feedback effects from electron-electron interactions on the momentum-space gap structure. We show how self-energy effects can lead to an orbital inversion of the orbital-resolved spin susceptibility, and thereby invert the hierarchy of the most important orbitals channels for superconducting pairing. This effect has important consequences for the detailed gap variations on the Fermi surface. We expect such self-energy feedback on the pairing gap to be generally relevant for superconductivity in strongly correlated multi-orbital systems. 

\end{abstract}
\maketitle

\section{Introduction} 

The discovery of a growing class of unconventional multi-band superconductors and the continuing development of high-resolution experimental probes of superconducting gap structures, highlight our need for theoretical models capable of quantitatively describing the detailed momentum structure of superconducting pairing. Thus, the modern theoretical objective is not only to locate the leading irreducible representation of the preferred gap function, but additionally to correctly describe all superconducting ``gap details'', consisting, for example, of the gap amplitude modulations and relative signs of the gaps on the participating Fermi surface sheets\cite{HirschfeldCRAS,ChubukovAR2012}. This ambitious goal has driven part of the research of iron-based superconductivity over the past decade, and remains currently very relevant, for example, for explaining superconductivity in multi-band compounds such as infinite layer nickelates\cite{Li2019}, FeSe\cite{Kreisel_review}, KFe$_2$As$_2$\cite{Okazaki2012}, LiFeAs\cite{Allan12}, UTe$_2$\cite{Jiao20}, and Sr$_2$RuO$_4$\cite{romer2019}. Since the superconducting gap structure is directly related to the mechanism of superconductivity, it serves as an important testing ground for different electron pairing scenarios. However, as exemplified in this paper, even within a given mechanism there are additional electron correlation effects that may severely affect the ``gap details'' of the resulting gap structure. 

The importance of the strength of Coulomb repulsion in the formation of Cooper pairs relates to the overall question of whether electron-electron interactions are friend or foe of superconductivity. Within most unconventional pairing scenarios, repulsive interactions generate pairing in the first place\cite{Kohn1965,Scalapino_RMP,Romer2020}, but there exists an optimal strength, since too strong interactions limit superconductivity. This issue has been discussed for cuprate superconductors where the single-band Hubbard Hamiltonian is the prime model under investigation\cite{Maier2006}. However, for multi-band systems there exist additional complexity since the states at the Fermi level may exhibit different orbital character\cite{Graser2009}. Thus, even though the Coulomb interaction is identical for electrons in the various active orbitals, the crystal structure and the corresponding band structure differentiates the orbitals at the Fermi level. At the bare level, one would expect the orbitals mainly contributing to Fermi level states to also dominate the pairing\cite{Kreisel2017}. However, Coulomb repulsion may cause these same dominant orbital states to exhibit the largest self-energy, with potentially important implications for the resulting pairing state\cite{Arakawa_2011,Yin2014,Yu2014,Nourafkan16,Kreisel2017,Hu2018,Lee_2018,Acharya_2019,Bhattacharyya_PRB2020_2,Fanfarillo_2020}.  

The issue of important self-energy feedback effects on the superconducting gap structure, has recently been intensely studied in relation to the material FeSe, featuring a Fermi surface consisting of small elliptical hole (electron) pockets near the $\Gamma$ ($M$) point of the 2-Fe Brillouin zone\cite{BoehmerKreisel_review,Coldea_2017_review,Kreisel_review}. In this compound, superconductivity materialises out of an electronic nematic phase and exhibits a remarkably large 90 degree rotational anisotropy\cite{Sprau2017}. The gap variation along the Fermi pockets is not straightforwardly explained by standard pairing models\cite{Sprau2017,Kreisel2017}. Specifically, the superconducting gap features very strong amplitude variations along the elliptical Fermi pockets and nearly vanishes around the vertex of the ellipses\cite{Sprau2017}. This gap structure is qualitatively different from that obtained from standard multi-orbital random phase approximation (RPA) spin-fluctuation pairing, an approach that has been generally successful for many other iron-based materials\cite{HirschfeldCRAS,Kuroki2008,Graser2009,Kreisel2017}. Therefore, it was hypothesized that additional correlation effects, e.g. orbital-dependent mass renormalizations, not included within the standard bare-RPA pairing formalism, might non-trivially affect the gap structure\cite{Kreisel2017,Benfatto2018,Hu2018,Rhodes_2020,Kang2018}.

Here, by performing a direct comparison of the superconducting gap structures arising from bare-RPA versus the self-consistent fluctuation exchange (FLEX) formalism\cite{Bickers_FLEX}, we can directly explore how electronic correlations can influence the resulting superconducting gap. Indeed, while both approaches subscribe to standard spin-fluctuation mediated superconductivity, the main difference between the methods is the inclusion of self-energy effects in the FLEX approach. While several earlier works have applied the self-consistent FLEX method to study pairing in iron-based superconductors\cite{Arita_Ikeda_2009,AritaIkeda2010,ikeda_2010}, none have focused on the detailed orbital contributions and contrasted these to simpler methods neglecting self-energy effects in the pairing kernel. We find that while the overall symmetry class of the leading gap solution is not affected (for our case study), there are significant ``gap details'' that get strongly modified by self-energy effects. We trace the main difference between bare-RPA and self-consistent FLEX to the presence of a so-called orbital inversion, i.e. the fact that the orbital structure of the pairing kernel has been restructured by self-energy effects. This change of the orbital hierarchy in the contributions to the pairing naturally arise in multi-orbital models where pairing and self-energy are caused by the same virtual processes, thereby rendering feedback effects important. We expect future computational schemes for pairing superior to FLEX to reach similar conclusions. 

Finally, we discuss an approximate method of including the self-energy feedback effects in the pairing by incorporating orbital-dependent quasiparticle weights as proposed in Refs.~\onlinecite{Kreisel2017,Sprau2017}.
We compare the results of this simpler approach to the full FLEX calculation.

\section{Model and Method} 

The model consists of a five-orbital tight-binding Hamiltonian $H_0$ including all relevant five Fe $d$ orbitals $ [ d_{x y},d_{x^2 - y^2}, d_{x z},d_{y z},d_{3z^2 -r^2} ]$, and onsite interactions via the standard Hubbard-Hund term $H_I$ 
    \begin{align} \label{Hamiltonian}
    \begin{split} 
     H   &= H_0 + H_I \\
        &= \sum_{ij\sigma} \sum_{qt}  t_{ij}^{tq} 
        c^\dagger_{it\sigma} c_{jq\sigma} \\
        & + U \sum_{it} n_{it\uparrow} n_{it\downarrow} + U' \sum_{i,t<q} \sum_{\sigma \sigma'} n_{it\sigma} n_{iq\sigma'} \\
        & + J \sum_{i,t<q} \sum_{\sigma \sigma'} c^\dagger_{it\sigma} c^\dagger_{iq\sigma'} c_{it\sigma'} c_{iq\sigma}  \\
        & + J' \sum_{i,t\neq q} c^\dagger_{it\uparrow} c^\dagger_{iq\downarrow} c_{it\downarrow} c_{iq\uparrow},
    \end{split}
    \end{align}
with interaction parameters $U,U',J,J'$ given in
the notation of Kuroki \textit{et al.}~\cite{Kuroki2008} fulfilling the relations $U'=U-2J$ and $J=J'$. Here $q$ and $t$ are orbital indices and $i$, $j$ denote Fe-atom sites. The kinetic part, $H_0$, is identical to that used in Ref. \onlinecite{Bhattacharyya_PRB2020_1} consisting of a DFT-derived five-band model generated for LiFeAs. We choose a band relevant for LiFeAs since this material does not exhibit magnetic and nematic instabilities, but stress that the discussion below is general, and should be of relevance also to other multi-band unconventional superconductors. The detailed band structure is not of crucial importance as long as it features multiple-orbital Fermi surface sheets.

The single-particle Green's function is given by
\begin{align} \label{Gfull}
    G(\mathbf{k},\omega_{m})^{-1} = G^0 (\mathbf{k},\omega_{m})^{-1} - \Sigma(\mathbf{k},\omega_m),
\end{align}
where the bare Green's function is $G^0(\mathbf{k},\omega_{m}) = \left[ i\omega_{m} - H_0(\mathbf{k}) + \mu\right]^{-1}$, and the self-energy in orbital basis is given by
    \begin{align}   \label{selfenergy}  
     \Sigma_{ps}(\mathbf{k},\omega_m)&\notag\ \\
    = \frac{1}{\beta N_q}&\displaystyle{\sum_{\mathbf{q},\Omega_{m}}\sum_{qt}} V_{pqst}(\mathbf{q},\Omega_{m}) G_{qt}(\mathbf{k-q},\omega_{m}-\Omega_{m}).
    \end{align}
Here, $V_{pqst}(\mathbf{q},\Omega_{m})$ refers to the effective particle-hole interaction given by
 \begin{align} \label{Veqn}
 \begin{split}
     V_{pqst}(\mathbf{q},\Omega_{m}) &= \left[ \frac{3}{2} U^S \chi^S(\mathbf{q},\Omega_{m}) U^S + \frac{1}{2} U^C \chi^C(\mathbf{q},\Omega_{m}) U^C \right. \\
     &\!\!\!\!\!- \left. \left(\frac{U^C+U^S}{2}\right) \chi^0(\mathbf{q},\Omega_{m}) \left(\frac{U^C+U^S}{2}\right) \right]_{pqst},
 \end{split}
 \end{align}
with the charge- and spin-fluctuation parts of the RPA susceptibility defined by
    \begin{align} \label{RPAsus}
    \begin{split} 
    \chi^C(\mathbf{q},\Omega_m) &=  [ 1 + \chi(\mathbf{q},\Omega_m) U^C]^{-1} \chi(\mathbf{q},\Omega_m) ,\\
    \chi^S(\mathbf{q},\Omega_m) &=  [ 1 - \chi(\mathbf{q},\Omega_m) U^S]^{-1} \chi(\mathbf{q},\Omega_m) ,
    \end{split}
    \end{align}
with the matrices $U^C$ and $U^S$ given by
   \begin{align} \label{U_matrix}
   \begin{array}{ll}
	 U^{C}_{pppp}= U ,    &  \quad U^{S}_{pppp}= U  \\
	 U^{C}_{ppss}= 2U'-J, &  \quad U^{S}_{ppss}= J  \\
	 U^{C}_{pssp}= J',    &  \quad U^{S}_{pssp}= J' \\
	 U^{C}_{psps}= 2J-U', &  \quad U^{S}_{psps}= U'. \\
   \end{array}
   \end{align}
The orbitally resolved susceptibility is given by
\begin{align} \label{baresus}
    & \chi_{pqst}(\mathbf{q},\Omega_m)\notag \\
    &= -\frac{1}{\beta N_k }  \sum_{\mathbf{k},\omega_m} G_{tq}(\mathbf{k},\omega_m) G_{ps}(\mathbf{k+q},\omega_m+\Omega_m)\,,
\end{align}
where $ N_k $ is the number of $\mathbf{k}$ points and $\beta=1/T$ denotes the inverse temperature. In Eq.(\ref{Veqn}), $ \chi^0(\mathbf{q},\Omega_{m})$ refers to the standard bare multi-orbital Lindhard function.

\begin{figure}[tb]
  \includegraphics[width=0.8\linewidth]{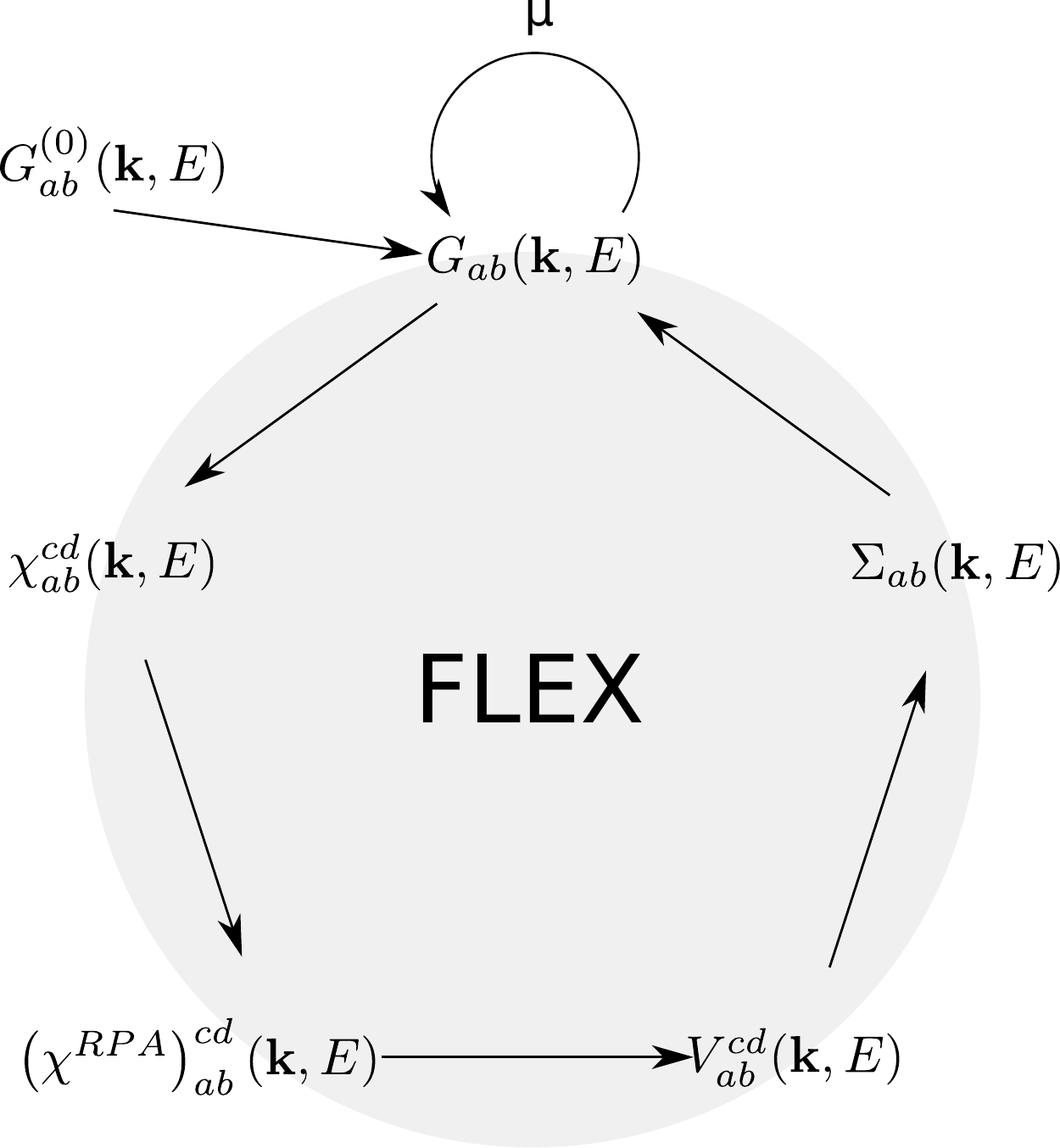}
  \caption{Schematic illustration of the self-consistent FLEX scheme.}
  \label{Figure:FLEX}
\end{figure}

To obtain a self-consistent solution, we solve Eqs. \eqref{Gfull}-\eqref{baresus} iteratively using the FLEX scheme shown in Fig.~\ref{Figure:FLEX}.
The circle at the top corresponds to a bisection search that is used to find a chemical potential $\mu$ such that the particle number is kept at $6.0$ within a numerical accuracy of $10^{-5}$.
To improve the numerical stability of the algorithm, the self-energy is also mixed with that obtained in the previous step using $\Sigma = \left(\Sigma_\text{new} + \Sigma_\text{previous}\right)/2$, where $\Sigma_\text{new}$ and $\Sigma_\text{previous}$ are the self-energies obtained in the current and previous steps, respectively.
The condition used to terminate the FLEX loop is that $\max(|G_\text{new} - G_\text{previous}|)/\max(|G_\text{previous}|) < 10^{-3}$, where $G_\text{new}$ and $G_\text{previous}$ are the Green's functions calculated in the current and previous steps, respectively. The calculations are done for $T=100\textrm{ K}$, using a momentum space mesh with $30\times 30$ points and truncating the Matsubara sums to include the 1000 smallest frequencies.

To compute the superconducting pair potential, we use the converged charge- and spin-susceptibilities to calculate~\cite{AritaIkeda2010}
\begin{align} \label{pairpotential}
 \begin{split}
     V_{pqst}^{(SC)}(\mathbf{q},\Omega_{m}) &= \hat{U} + \frac{3}{2} U^S \chi^S(\mathbf{q},\Omega_{m}) U^S\\
     &- \frac{1}{2} U^C \chi^C(\mathbf{q},\Omega_{m}) U^C,
 \end{split}
 \end{align}
 where
\begin{align} \label{U_hat_matrix}
   \begin{array}{ll}
	 \hat{U}_{pppp}= U,\quad
	 \hat{U}_{ppss}= U',\\
	 \hat{U}_{pssp}= J,\quad
	 \hat{U}_{psps}= J'.\\
   \end{array}
\end{align}

We also obtained the spectral function through analytical continuation of the converged Green's function using the Pad\'e approximation.
More specifically, we use the continued fraction method~\cite{Vidberg1977,Beach_2000}.

The appearance of a superconducting instability at $T_c$ can be found from solving the linearized gap equation,

 \begin{equation}\label{eqn:gapeqn}
-\frac{1}{V_G}  \sum_\mu\int_{\text{FS}_\mu}dS'\; \Gamma_{\nu\mu}(\mathbf{k},\mathbf{k}') \frac{ g_i(\mathbf{k}')}{|v_{\text{F}\mu}(\mathbf{k}')|}=\lambda_i g_{i}(\mathbf{k})\,,
 \end{equation}
 for the eigenvalues $\lambda_i$ and the eigenvectors $g_i(\mathbf{k})$. Here $\mu,\nu$ are the band indices of the Fermi surface vectors $\mathbf{k},\mathbf{k}'$, respectively, $V_{G}$ is the area of a Brillouin zone and the magnitude of the Fermi velocity $|v_{\text{F}\mu}(\mathbf{k})|$ weights the corresponding Fermi point.
 For this purpose, we project the pairing interaction at the lowest frequency into band space by
 \begin{align}
	&{\Gamma}_{\nu\mu} (\mathbf{k},\mathbf{k}')  = \mathrm{Re}\sum_{pqst} a_{\nu}^{p,*}(\mathbf{k}) a_{\nu}^{t,*}(-\mathbf{k}) \nonumber\\&\hspace{0.6cm}\times 
	{V}^{(SC)}_{pqst} (\mathbf{k}-\mathbf{k}',\Omega_1) \;  a_{\mu}^{q}(\mathbf{k}') a_{\mu}^{s}(-\mathbf{k}')\, \label{eq_Gam_mu_nu}\,,
\end{align}
where $a_{\mu}^{p}(\mathbf{k})$ are the matrix elements from the transformation from orbital space to band space. Here, we use as Hamiltonian $\tilde H(\mathbf{k})=H_0(\mathbf{k})+\frac 12 [\Sigma(\mathbf{k},\omega_0)+\Sigma(\mathbf{k},\omega_0)^\dagger]$ where the Hermitian part of the self-energy matrix $\Sigma(\mathbf{k},\omega_0)$ at the lowest Matsubara frequency $\omega_0$, Eq. (\ref{selfenergy}) is considered, a very good approximation at the Fermi level.
The leading instability, identified by the largest eigenvalue $\lambda_i$ is strictly speaking only correctly identified exactly at the critical temperature $T_c$, thus also the superconducting order parameter $\Delta(\mathbf{k})$ is proportional to $g_i(\mathbf{k})$ in this regime. We contrast this calculation to two usual spin-fluctuation calculations where the effective interaction is  obtained within RPA without self-energy corrections~\cite{Graser2009}, and an approach where the effect of correlations is included via quasiparticle weights in the susceptibility that enters RPA and the projection to band space given in Eq. (\ref{eq_Gam_mu_nu}) \cite{Kreisel2017}.

\begin{figure*}[tb]
  \includegraphics[width=\linewidth]{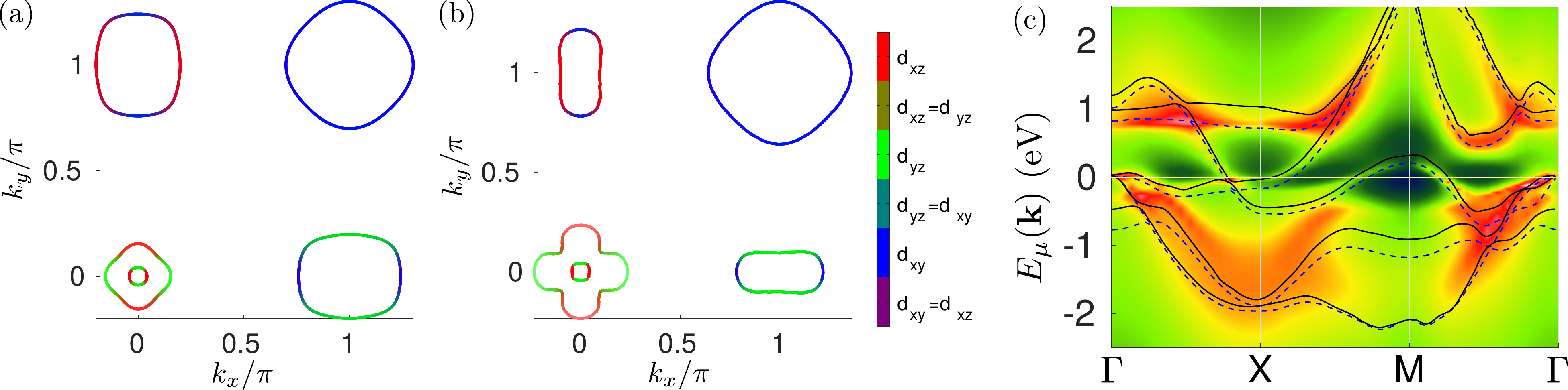}
  \caption{(a) Fermi surface of the bare model. (b) Fermi surface of the correlated model with self-energy corrections and (c) Spectral function (colormap) from the converged FLEX calculation for the same parameters as in (b), together with the unrenormalized bands (blue dashed) and the bands as obtained from adding the Hermitian part of the self energy (black).
  \label{fig_fermi_compare}
  }
\end{figure*}

\section{Results} 

The role of electron correlations on the band structure of FeSCs has been studied theoretically by application of a variety of different methods \cite{Miyake_cRPA,Medici_review,Biermann_review,Bascones_review}. In particular, both local and non-local self-energy effects on the Fermi surface have been explored and compared to experiments, and LiFeAs has played a prominent role in this exploration\cite{Yin2011_mag,Platt_2011,Ferber_2012,Wang13,Saito_LiFeAS_SOC,Saito14,Ahn14,Zantout_2019,Bhattacharyya_PRB2020_1,Kim_2020,Scherer2017Mar}.  
Figure \ref{fig_fermi_compare} compares the bare electronic structure to the one described by the FLEX approach including the self-energy from Eq.~(\ref{selfenergy}) with $U=1.2$ eV, $J=0.15$ eV. In Fig. \ref{fig_fermi_compare}(c) we show the associated low-energy band structure along a standard high-symmetry momentum cut, again comparing the bare eigenenergies to the spectral density as extracted from the total spectral function continued to the real axis by use of the Pad\'{e} approximation. Figure \ref{fig_fermi_compare}(a) displays the bare Fermi surface, whereas Fig. \ref{fig_fermi_compare}(b) shows the Fermi surface as obtained by utilizing the approximate Hamiltonian $\tilde H(\mathbf{k})$. In Fig. \ref{fig_fermi_compare}(c) the physical spectral density is compared to the band structure arising from both $H_0({\mathbf{k}})$ and $\tilde H(\mathbf{k})$. As seen, the latter exhibits good overall agreement with high spectral intensity at low energies. From Fig.~\ref{fig_fermi_compare} it is evident that the main effect of $\Sigma_{ps}(\mathbf{k},\omega_m)$ is to slightly modify the Fermi surface, causing smaller (larger) electron (hole) pockets. In addition, the second hole pocket around $\Gamma$ acquires a ``flower'' shape. Such Fermi surface changes due to a momentum-dependent self-energy have been investigated in detail previously \cite{ikeda_2010,Zantout_2019,Ortenzi2008,Arita_Ikeda_2009,Bhattacharyya_PRB2020_2}, and will not be further elaborated here. Instead, we focus on the modifications of the pairing structure from interactions.

\begin{figure*}[t]
  \includegraphics[width=\linewidth]{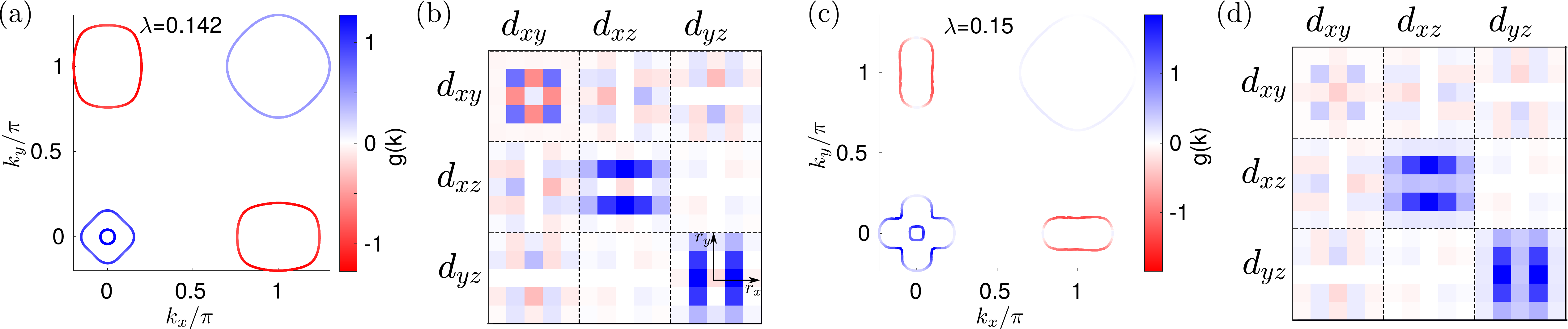}\caption{
  (a) Superconducting gap symmetry function $g({\mathbf k})$ for the pairing calculation in RPA ($U=0.9\,\mathrm{eV}$, $J/U=1/8$) together with (b) the transformation to real space pairing amplitudes and projection to orbital space. (c) the corresponding calculation using the pairing interaction as obtained from the self-consistent FLEX approach, $U=1.2 \,\mathrm{eV}, J/U=1/8$. The latter yielding a much smaller pairing amplitude at the $k$ points dominated by the $d_{xy}$ orbital which is also visible in the much smaller pairing amplitudes in the $d_{xy}$-$d_{xy}$ channel (first square) of panel (d).
  \label{fig_gap_compare} }
\end{figure*}

Thus, we turn to a discussion of the superconducting gap structure, contrasting the bare RPA case to the FLEX situation including self-energy corrections. In the RPA calculation, the interaction parameters have to be considered as effective values, leading to a nominally smaller critical value for the Stoner instability. We therefore adjust the value of $U$ such that the resulting eigenvalues $\lambda$ in the pairing calculations are almost identical, leading to $U=0.9\,\mathrm{eV}$ for the RPA case. In Fig. \ref{fig_gap_compare} we display both the gap along the Fermi surface pockets \ref{fig_gap_compare}(a,c) and the real-space orbital structure of the resulting gap function \ref{fig_gap_compare}(b,d). As seen, both gap structures support an $s\pm$ superconducting gap solution. However, there are significant effects of the self-energy on the ``gap details''. First of all, as seen from both the momentum-space plots and the real-space plots, the $d_{xy}$ orbital contribution is significantly reduced in the FLEX case. This is evident from comparison of the relative strength of the intra-orbital pairing seen in the top left square of Fig. \ref{fig_gap_compare}(b,d). In addition, this reduction is evident from the very weakened gap amplitude in $d_{xy}$-dominated segments of the Fermi surface, as seen from comparing Fig. \ref{fig_gap_compare}(a,c). Essentially the blue sections of the Fermi surface shown in Fig. \ref{fig_fermi_compare}(b) have been  ``washed out''. 

In band space, the hierarchy of the gaps remains the same, e.g. the inner $\Gamma$-centered hole pocket exhibits the largest gap in both Fig. \ref{fig_fermi_compare}(a,c). The reduction of $d_{xy}$-dominated pairing renders the gap at the $M$ point very small in the FLEX case, and it boosts the gap amplitude modulation on pockets exhibiting mixed orbital content, particularly those containing $d_{xy}$-dominated sections. This is seen most clearly from the electron pockets centered at the $X$ and $Y$ points in Fig. \ref{fig_gap_compare}(c).
We additionally show the gap structure projected to orbital space and Fourier transformed to real space,  Fig. \ref{fig_gap_compare}(b,d) for the two calculations. Each block represents the pairing in a certain orbital component $\Delta_{ps}(\mathbf{r})$, where the onsite pairing $\mathbf{r}=0$ is represented by the center square of each block, and bond order parameters by the appropriate squares away from the center. Indeed, one can note that the $d_{xy}-d_{xy}$ block is strongly suppressed for the FLEX calculation (panel (d)) such that the $d_{xz}-d_{xz}$ and $d_{yz}-d_{yz}$ components dominate the FLEX result.

As a particular feature, we point out the structure of two blocks in the pairing field for the conventional calculation, Fig.~\ref{fig_gap_compare}(b):  In the $d_{xz}-d_{xz}$ [$d_{yz}-d_{yz}$] block, the $\mathbf{r}=(\pm 1,0)$ [$\mathbf{r}=(0,\pm 1)$] bond order $\Delta_{xz,xz}(\pm 1,0)$ [$\Delta_{yz,yz}(0, \pm 1,0)$] is small, but has opposite sign compared to the $\mathbf{r}=(0,\pm 1)$ [$\mathbf{r}=(\pm 1,0)$] bond order $\Delta_{xz,xz}(0,\pm 1)$ [$\Delta_{yz,yz}(\pm 1,0)$]. Formally, the nearest-neighbor pairing in the $d_{xz}$ [$d_{yz}$] orbitals can be decomposed into an $s$-wave contribution, $\tilde\Delta_s$, and a $d$-wave contribution,  $\tilde\Delta_d$, such that $\Delta_{xz,xz}(\pm 1,0)=\tilde\Delta_s-\tilde\Delta_d\approx -0.001$ and $\Delta_{xz,xz}(0,\pm 1)=\tilde\Delta_s+\tilde\Delta_d\approx 1.2$ [$\Delta_{yz,yz}(\pm 1,0)=\tilde\Delta_s+\tilde\Delta_d$ and $\Delta_{yz,yz}(0,\pm 1)=\tilde\Delta_s-\tilde\Delta_d$]. The sign change of the bond order then implies that 
$\tilde\Delta_d$ is slightly larger than $\tilde\Delta_s$. 
The order parameter $\tilde\Delta_d$ encodes an orbital singlet pairing with $d_{x^2-y^2}$-form factor as for example worked out in Refs. \cite{Hu2018,Nica_2020} for a different compound. In the longer range pairing channels like e.g. $\Delta_{xz,xz}(0,\pm 2)$ and $\Delta_{yz,yz}(0,\pm 2)$ the change of sign is more visible, even though the overall size of the pairing strength at these sites is much smaller compared to the nearest neighbor pairing strength.
The orbital singlet nature of the gaps is suppressed in the FLEX calculation, as seen from Fig.~\ref{fig_gap_compare}(d). Here, all pairing amplitudes in the $xz-xz$ and $yz-yz$ blocks are positive (blue). This means that the orbital character of the gap is preferentially ``triplet-like'' although a small singlet component remains, as seen from the change in gap magnitude between e.g. $\Delta_{xz,xz}(0,\pm 1)$ and $\Delta_{xz,xz}(\pm 1,0)$.

\begin{figure*}[tb]
  \includegraphics[width=\linewidth]{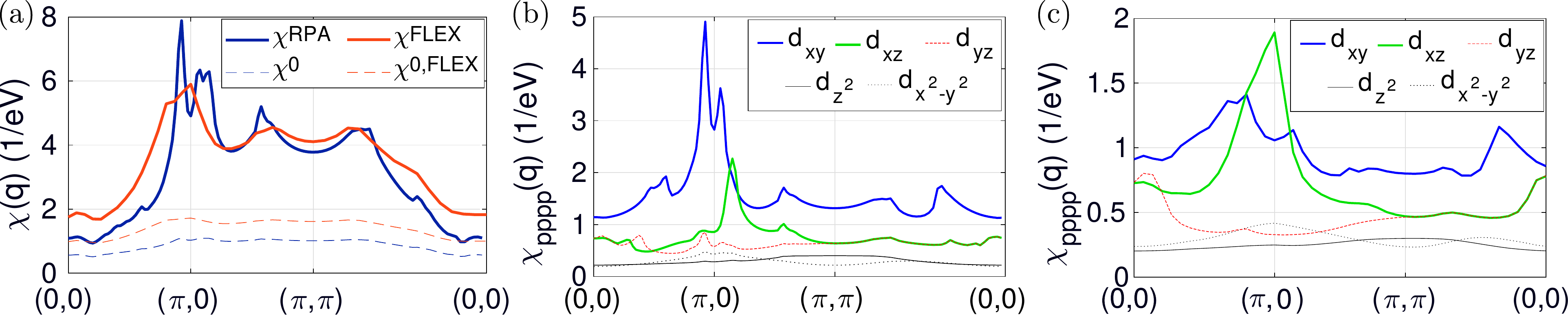}\caption{
  (a) Comparison of the physical susceptibility as obtained from a non-self-consistent RPA calculation ($U=0.86$ eV) and the self-consistent FLEX approach, $U=1.2$ eV, $J=0.15$ eV. (b) orbitally resolved susceptibility of the non-self-consistent approach showing the dominant $d_{xy}$ contribution that gives rise to the peak at $(\pi,0)$ and (c) corresponding orbitally resolved susceptibilities from the FLEX approach with dominant $d_{yz}$ contribution at the same momentum transfer of $(\pi,0)$.\label{fig_chi_compare}}
\end{figure*}

What is the underlying reason for the modified superconducting gap structures seen from the comparison between Fig. \ref{fig_gap_compare}(a,c)? In order to answer this question we need to dissect the pairing kernel giving rise to the Cooper pair structure. As evident from Eqs. (\ref{pairpotential})-(\ref{eq_Gam_mu_nu}), the most important ingredient is the spin susceptibility. In Fig. \ref{fig_chi_compare}(a) we display the real-part of the static total spin susceptibility for the bare, RPA, and FLEX cases. As seen, the overall momentum structure is preserved, and FLEX merely tends to smear out the peak structure as a direct result of the self-energy. However, as seen from a comparison of the orbital-resolved susceptibilities in Fig. \ref{fig_chi_compare}(b,c), an orbital inversion has taken place near $X=(\pi,0)$ (and similarly near $Y$, not shown). In the FLEX case, the dominant susceptibility is found in the $d_{xz}/d_{yz}$ channel, as opposed to the $d_{xy}$ orbital channel for the bare-RPA case, thereby fundamentally restructuring the dominant pair scattering contributions to superconducting pairing. This orbital inversion of the susceptibility components is the main reason for the differences between the bare-RPA and FLEX superconducting gaps. The orbital inversion arises quite naturally in FLEX, since the same physical processes are involved in pairing and the self-energy. Thus, for systems close to magnetic instabilities where the spin susceptibility (bubble and ladder diagrams) is expected to be particularly important, incorporating the self-energy feedback effects of the susceptibility is crucial for obtaining a correct description of the ``gap details''. At present it remains to be seen whether future experiments, e.g. RIXS, may be able to selectively probe the orbital content of the spin susceptibility and compare this to calculations based on the bare band structure to verify the existence of such self-energy-generated orbital inversion. 

As a function of increased interaction parameters, $U$, $J$, both RPA and the FLEX approximation exhibit an instability of a magnetic state. In general, the self-energy in the FLEX approach tends to broaden the peaks of the susceptibility, thereby pushing the Stoner instability to a larger critical interaction strength. The orbital inversion, and the concomitant effects on the superconducting gap structure, discussed above are found in this enlarged large-$U$ regime, and the results presented here are robust in this regime. At weak interaction strengths, the FLEX results resemble the bare-RPA results. Finally, we stress that for other bands the self-energy effects may not be as simple as an orbital inversion, especially if the momentum positions of the dominant peaks in the susceptibilities shift.

The result discussed above, exhibiting a strongly suppressed gap on the $d_{xy}$-dominated Fermi surface sections, is an example of what has been dubbed orbital-selective superconductivity.  A ``poor man's'' version  to incorporate
 the self-energy effects has been proposed e.g. in Refs. \onlinecite{Sprau2017,Kreisel2017,Zhou2020, kreisel2018}
 applied to FeSe and also in Ref. \onlinecite{romer2019} applied to Sr$_2$RuO$_4$. In this method, quasiparticle weight renormalizations are simply included at the level of the single-particle Green's function. In this way, orbitals expected to be exposed to the largest self-energy effects, thereby experiencing the most suppressed quasiparticle weights, are naturally suppressed in their contributions to the superconducting pairing kernel. 

We demonstrate this effect in the following; we can extract the approximate FLEX quasiparticle weights directly from the calculated self-energy by $Z_\alpha(\mathbf{k})\approx \left[1- {\mathrm{Im}} ~\Sigma_{\alpha\alpha}(\mathbf{k},\omega_0)/\omega_0 \right]^{-1}$, with $\omega_0=\pi T$\cite{AritaIkeda2010,ikeda_2010}. This procedure yields $Z_{xy}\approx 0.52$ and $Z_{xz}=Z_{yz}\approx 0.6$ when evaluated at their relevant Fermi surface points, for the current FLEX case with $U=1.2$ eV, $J=0.15$ eV.  Note that the relevant quasiparticle weights are only those for the $d_{xy}, d_{xz}, d_{yz}$ orbitals. The two other orbitals have negligible weight on the Fermi surface and therefore their values do not influence the result for the pairing\cite{Kreisel2017} and these orbitals contribute much less to the susceptibility, see Fig. \ref{fig_chi_compare} (b). For the case of FeSCs, it is well-known from a wide range of methods that indeed $Z_{xy}$ exhibits the largest reduction compared to the other four 3d orbitals\cite{Medici_review,Biermann_review,Yin2011_mag,Bascones_review,Yu_Si_2012,Fanfarillo2017}. In terms of the FLEX method, the fact that the $d_{xy}$ orbital exhibits the largest mass renormalization for FeSCs was studied in detail e.g. in Refs. \onlinecite{AritaIkeda2010,ikeda_2010,Bhattacharyya_PRB2020_1}. 

\begin{figure}[tb]
  \includegraphics[width=\linewidth]{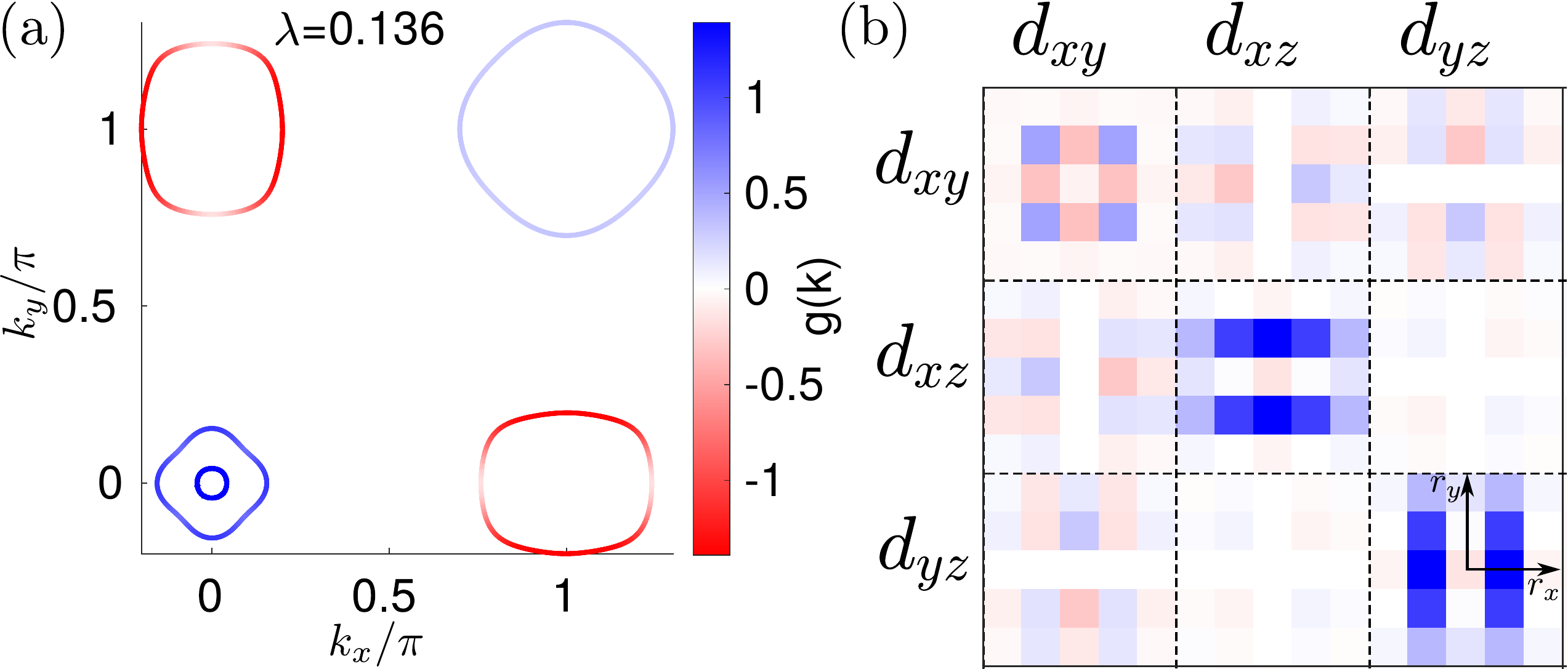}\caption{
  (a) Superconducting gap symmetry function $g({\mathbf k})$ for the pairing calculation using a modified spin-fluctuation pairing calculation with $U=2.5$ eV,  $J/U=1/8$, and quasiparticle weights $Z_{xy}=0.52$, $Z_{xz/yz}=0.6$, $Z_{x^2-y^2}=0.8$ and $Z_{z^2}=0.72$ as deduced from the FLEX calculation. (b) Corresponding real-space pairing amplitudes showing small pairing contributions in the $d_{xy}$ channel, while preserving the sign change in the next-nearest-neighbor pairing amplitudes of the $d_{xz}$ and $d_{yz}$ orbital components, i.e. $\Delta_{xz,xz}(0,\pm 2)-\Delta_{xz,xz}(\pm 2,0)=-[\Delta_{yz,yz}(0,\pm 2)-\Delta_{yz,yz}(\pm 2,0)]$.
  \label{fig_gap_orb_sel} }
\end{figure}

Using the above-extracted values for the quasiparticle weights to renormalize the susceptibility and projection matrices of the usual RPA method leads to the superconducting gap structure shown in Fig. \ref{fig_gap_orb_sel} where we have chosen the effective interaction ($U=2.5 \,\mathrm{eV}$) such that, again, the eigenvalue for the pairing calculation matches the one from the previous FLEX calculation; this choice, however, does not alter the result of the order parameter $g(\mathbf{k})$. As seen, including a reduced weight on the $d_{xy}$ orbital qualitatively reproduces the FLEX very well (except for some details, e.g. the orbital-singlet pairing amplitude is not appropriately suppressed, $\tilde\Delta_d>\tilde\Delta_s$).

\section{discussion and conclusions} 

The work presented here is related to another recent theoretical study of momentum-dependent quasiparticle renormalization on the gap structure of FeSCs\cite{Bhattacharyya_PRB2020_2}. In Ref.~\onlinecite{Bhattacharyya_PRB2020_2} a one-loop FLEX calculation was presented, and it was explored how the inclusion of quasiparticle renormalizations $Z({\mathbf{k}})$ alter the ``gap details'' along the relevant Fermi surface sheets. In agreement with the current work, the $d_{xy}$ dominated part of the electron pockets were suppressed compared to the bare-RPA approach. However, no gap reduction was found in Ref.~\onlinecite{Bhattacharyya_PRB2020_2} on the large $d_{xy}$-dominated hole pocket centered at $M$, as opposed to the findings reported in this work. The main methodological difference with respect to the current work is the inclusion of the full self-consistency loop incorporated here. This allows for 1) proper self-energy feedback effects on the susceptibilities entering the pairing kernel, and 2) exploration of the regime of larger interactions $U$, $J$ inaccessible to one-shot calculations. These properties are crucial for the orbital inversion discovered here, and cause the strongly suppressed pairing on all $d_{xy}$-dominated sections of the Fermi surface.

More generally, we expect that other strongly correlated multi-orbital metals should exhibit similar self-energy effects of their gap structure. However, there may well also be other cases where the self-energy can severely alter the gap for other reasons, e.g. by restructuring the orbital content of the Fermi surface, or by fundamentally altering the dominant momentum structure of the spin susceptibility. In the latter case, one can expect that self-energy effects change the symmetry class of the leading pairing instability \cite{Bhattacharyya_PRB2020_2}.  

In summary, we have explored the role of electronic correlations on the superconducting gap structure in multi-orbital systems. Specifically, we contrasted the gap structure obtained from standard bare-RPA with that generated within the FLEX formalism, including self-energy feedback effects on the superconducting pairing kernel. The main finding is the existence of an orbital inversion, caused by the self-energy, of the hierarchy of susceptibility channels contributing to the pairing. This reduces the gap on the strongest correlated orbitals. For the current band structure, this leads to a very small gap on the largest hole pocket, and a significantly enhanced gap amplitude modulation on the electron pockets. This demonstrates the relevance of self-energy feedback effects on the gap structure of multi-orbital correlated metals.\\

\section*{Acknowledgements} 
We thank T. A. Maier and P. J. Hirschfeld for useful conversations. K. B. and B. M. A. acknowledge support from the Independent Research Fund Denmark grant number 8021-00047B.

%

\end{document}